\documentclass{article}
\usepackage{graphicx} 
\usepackage{bbm}
\usepackage{amsfonts}
\usepackage{amssymb}
\usepackage{amsmath}

\newcommand{\ii}{\mathsf{i}}

\newcommand{\bu}{\mathbf{u}}

\newcommand{\R}{\mathbb{R}}

\newcommand{\C}{\mathbb{C}}

\def\lg{\langle }
\newcommand{\rg}{\rangle }

\newcommand{\ud}{\mathrm{d}}

\newcommand{\bun}{\mathbbm{1}}

\def\upa{| \uparrow\,\rg}
\def\dwa{| \downarrow\,\rg}

\title{The Geometry of Qubit Decoherence:\\ Linear vs. Nonlinear Dynamics in the Bloch Ball}
\author{Alan C. Maioli $^{1}$, Evaldo M. F. Curado $^{1,2}$, Jean-Pierre Gazeau$^{3,4}$ and Tomoi Koide$^{5}$}
\date{October 2025}

\begin{document}

\maketitle
$^{1}$ \quad Centro Brasileiro de Pesquisas F\'{\i}sicas\\
$^{2}$ \quad National Institute of Science and Technology for Complex Systems\\
\ \ \ \quad Rua Xavier Sigaud 150,  Rio de Janeiro, Brazil\\
$^{3}$\quad Université Paris Cité, CNRS, Astroparticule et Cosmologie, 75013 Paris, France \\ 
$^{4}$\quad Faculty of Mathematics, University  of Bia\l ystok, 15-245 Bia\l ystok, Poland \\
$^{4}$ \quad Instituto de Física, Universidade Federal do Rio de Janeiro, C.P. 68528, 21941-972, Rio de Janeiro, RJ, Brazil
\begin{abstract}
   We present two complementary approaches to the GKSL equation for an open qubit. The first, based on linearity, yields solutions illustrated by mixed states trajectories 
 in the Bloch ball, including non-random asymptotic  fixed points, and exceptional points. The second, exploiting the SU$(2)$ symmetry, leads to a nonlinear dynamical system that separates angular dynamics from radial dissipation. This symmetry-based perspective offers a promising route toward generalisation to open qudits.
\end{abstract}

\section{Introduction}

Studies in quantum physics have traditionally focused on systems isolated from their environment.
However, with the recent developments in quantum computing \cite{Nielsenbook} and quantum thermodynamics \cite{opensystembook}, the dynamics of open quantum systems has attracted growing interest.
This line of research is also deeply connected to the measurement problem \cite{Dodonov2005}, an unresolved issue since the dawn of quantum mechanics, making it critically important from both fundamental and applied perspectives \cite{Alipour2014,Jirari2005,CircuitComplexity}.

The dynamics of such open quantum systems should, in principle, be derived
from the microscopic unitary evolution of the total system (system plus environment),
but this is generally difficult to perform. An axiomatic approach,
therefore, becomes effective. Specifically, we assume that the dynamics of the system of interest is linear and Markovian.
Under these assumptions, if we impose the requirement that the time evolution must preserve physical states,
that is, a complete positiveness and trace-preservation (CPTP) map,
it is known that the quantum master equation is
the Gorini-Kossakowski-Sudarshan-Lindblad (GKSL) equation \cite{opensystembook, rivas2012, shortintro, LindbladHistory}.
This equation unifies the treatment of the unitary evolution governed by the  Hamiltonian of the system and the non-unitary dissipative processes described by GLKS jump operators, which represent the interaction with the environment.
In particular, when applied to a single qubit (a two-level system), the fundamental unit of quantum information, its time evolution can be intuitively understood through a geometric picture using the Bloch vector \cite{BlochLindblad}.
In the representation based on Cartesian coordinates, this GKSL equation reduces to a tractable system of linear equations, and its solutions and properties have already been  studied (see for instance Chapt. 9 in  \cite{klimov09}). 
In the first part of our work, we revisit this system to analyze specific anisotropic cases, thereby clarifying how the interplay between Hamiltonian evolution and environmental dissipation gives rise to phenomena such as damped oscillations and the emergence of non-random, asymptotically stable fixed points within the mixed-state space. Our investigation highlights an important feature for the design of robust qubits: by engineering the qubit–environment interaction, one can protect selected quantum states \cite{Lidar,Pastawski,Verstraete,Altafini}. Furthermore, we explicitly identify the critical transition between oscillatory and monotonic decay regimes as an \emph{exceptional point} in the spectrum of the Liouvillian superoperator. This result establishes a direct link between the geometric coalescence of state trajectories within the Bloch ball and the spectral degeneracies that characterize non-Hermitian quantum mechanics.

In the second part of the article, we develop an alternative descriptive framework for the same physical system by reformulating the dynamics in spherical coordinates, a representation that offers additional geometric clarity and facilitates the interpretation of anisotropic dissipation, fixed-point structure, and trajectory behavior within the Bloch ball.
This coordinate transformation renders the equations of motion nonlinear, but it offers advantages that more than compensate for this complexity.
First, it reveals a clear variable separation structure. Specifically, the radial variable, which is related to the change in the system's purity (and thus the informational entropy due to interaction with the environment), is naturally decoupled from the angular variables, which describe the orientation of the quantum state.  This structure of variable separation is expected to hold universally, not only for two-level systems but also for general $N$-level systems.
This provides an insightful theoretical framework for uniformly treating the dynamics of more complex quantum systems.
Hence, we believe that our approach offers a new perspective on the geometric interpretation of the quantum information space. Also,  our unified geometric
presentation provides a pedagogically valuable perspective and a useful tool
for generalizations, because it combines linear and nonlinear perspectives, revealing complementary physical insights. Our approach goes beyond the standard linear solutions by deriving the explicit nonlinear equations of motion for the qubit's orientation, which we identify as having a Ricatti-like structure \cite{CircuitComplexity}. Additionally, we obtain analytical trajectory invariants depicted as constants of motion for the dissipative path. To the best of our knowledge, such closed-form geometric constraints for anisotropic GKSL evolution have not been reported in standard literature. The trajectory invariants provide a rigorous geometric description of the state's spiral towards the fixed point. This is followed by a discussion over the possible fixed points, which lie on the axes of the Bloch sphere representation.

The organisation  of the paper is as follows. In Section \ref{sec:GKSL} we introduce the GKSL equation in the qubit case and proceed with its standard linear analysis in Cartesian coordinates. We then study the isotropic case and particular anisotropic situations. The examples are illustrated by figures showing trajectories in the Bloch ball. In Section \ref{SU2open} we proceed with a SU$(2)$ approach to the GLKS system which leads to the nonlinear dynamical system in spherical coordinates whose the radial part shows the dissipative regime in a straightforward way. In Section \ref{fixed}, we carry out a detailed study of the existence of fixed points and explore how they can be exploited to control the evolution of a qubit.  In Section \ref{conclu} we give insights about possible generalizations. Some details of our computations are made explicit in Appendices \ref{ap: longtimes} and \ref{ap: DynSys}.    

\section{GKSL}\label{sec:GKSL}
  In this section, we review the standard formulation of the GKSL equation for a single qubit. We employ the Bloch vector representation, which maps the density matrix to a vector in $\mathbb{R}^3$, and derive the linear system of differential equations that governs its dynamics. The starting point is the GKSL equation in the Schr\"{o}dinger  picture for a density matrix (i.e., quantum state) $\rho$, with the standard sign convention: 
\begin{equation}
\label{eq: lindbladA}
\begin{split}
\frac{\mathrm{d} \rho}{\mathrm{d} t}&=
-\ii\,[H,\rho]+\sum _{k}h_k\left[L_{k}\rho L_{k}^{\dagger } -\frac{1}{2}\left(\rho L_{k}^{\dagger}L_{k}+L_{k}^{\dagger}L_{k}\rho\right)\right] \\
    &\equiv \mathcal{L}(\rho) = \mathcal{L}_{\mathrm{un}}(\rho)+\mathcal{L}_{\mathrm{diss}}(\rho)\,,
\end{split}
\end{equation}
where the  $L_k$'s together with the identity form an arbitrary basis of matrix operators and the coefficients $h_k$ are non-negative constants, and we set $\hbar=1$. We also distinguish between the unitary Hamiltonian regime ($\mathcal{L}_{\mathrm{un}}$) and the dissipative regime ($\mathcal{L}_{\mathrm{diss}}$) in the action of the Lindblad operator $\mathcal{L}$. In the present two-dimensional case, a natural basis would be  the set compounded of the identity $\bun_2$, the three Hermitian Pauli matrices $\sigma_i$, and the three antihermitian matrices $\ii\sigma_i$. Actually one shows that we just need the subset   $\{\bun_2, \sigma_i\, , \, i=1,2,3 \, \equiv x,y,z\}$.

Hence, writing  the density (i.e. (nonnegative hermitian unit trace) operator as 
\begin{equation}
    \rho=\frac{1}{2}\left(\mathbbm{1}_2+a_1 \sigma_1+a_2 \sigma_2+a_3 \sigma_3\right)\,,
\end{equation}
with the restriction $\sqrt{a_1^2+a_2^2+a_3^2}\leq 1$, the $L_k$ as the Pauli matrices, and the Hamiltonian written in Pauli's matrices basis,
\begin{eqnarray}\label{eq: HamilCartesian}
    H=e_0\mathbbm{1}_2+e_1 \sigma_1+e_2 \sigma_2+e_3 \sigma_3\,,
\end{eqnarray}
the GKSL equation results in the set of 3 equations (see appendix),
\begin{equation}\label{eq: dynsys1qbit}
    \begin{split}
        \dot{a}_1&=2(e_2a_3-e_3a_2)-2(h_2+h_3)a_1 \\
        \dot{a}_2&=2(e_3a_1-e_1a_3)-2(h_1+h_3)a_2 \\
        \dot{a}_3&=2(e_1a_2-e_2a_1)-2(h_1+h_2)a_3\,.
    \end{split}
\end{equation}
Thus, it can be rewritten in the following vector form characterizing a dynamical system,
\begin{equation}\label{eq: CartesianEvo}
\vec{\dot{a}}=2\vec{e}\times\vec{a}-2D_r\cdot\vec{a}\,,
\end{equation}
where $\vec{a}$ is the usual Bloch vector represented by a column matrix $\vec{a}=(a_1,a_2,a_3)^T$, $\vec{e}=(e_1,e_2,e_3)^T$, while $\times$ depicts the cross product. The matrix
\begin{equation}
    D_r=\begin{pmatrix}
        h_2+h_3 &0&0\\
        0&h_1+h_3&0\\
        0&0&h_1+h_2
    \end{pmatrix}\,,
\end{equation}
contains the decay rates $h_1$, $h_2$, and $h_3$. Additionally, we  write the cross product $\vec{e}\times\vec{a}$ as the product between the skew-symmetric matrix
\begin{equation}
    e_\times=\begin{pmatrix}
        0 & -e_3 & e_2\\
        e_3 & 0 & -e_1\\
        -e_2 & e_1 &0
    \end{pmatrix}\,,
\end{equation}
and vector $\vec{a}$. 
We  observe that the time derivative of the Bloch vector is equivalent to the matrix multiplication
\begin{equation}\label{eq: DynSystemSimpleForm}
\vec{\dot{a}}= T_d\cdot \vec{a}\,,
\end{equation}
where
\begin{equation}
    T_d=-2\begin{pmatrix}
        (h_2+h_3) & e_3 & -e_2\\
       - e_3 & (h_1+h_3) & e_1\\
        e_2 & -e_1 & (h_1+h_2)
    \end{pmatrix}\,.
\end{equation}
\subsection{Solutions to the system of ode: isotropic case}
In the case of time independent dissipative factors $h_k$, Equation \eqref{eq: DynSystemSimpleForm} can be solved by the well-known ansatz $\vec{a}(t) = \vec{v} e^{\lambda t}$, where $\vec{v}$ is a constant vector. Then, it becomes
\begin{equation}
    \lambda \vec{v}e^{\lambda t}= T_d \cdot \vec{v} \,e^{\lambda t}\,,
\end{equation} 
and the exponentials cancel out, leaving us with an eigenvalue problem. To simplify the formulation, we  consider the isotropic case $h_1=h_2=h_3=h$, which gives the following results for the eigenvalue problem:
\begin{align}
    \vec{v}_1 = (e_1,e_2,e_3)^T = \vec{e}\,, &\quad \lambda_1 = -4h\,, \\
    \vec{v}_2 = \left(e_1e_2 + \ii e_3 \Vert \vec{e}\Vert\,, \ -(e_1^2+e_3^2), \ e_2e_3-\ii e_1\Vert \vec{e}\Vert \right)^T\,, &\quad \lambda_2= -4h-2\ii\Vert \vec{e}\Vert\,, \\
    \vec{v}_3 = \left(-e_1e_2 + \ii e_3 \Vert \vec{e}\Vert\,, \ (e_1^2+e_3^2), \ -e_2e_3 -\ii e_1\Vert \vec{e}\Vert \right)^T\,, &\quad \lambda_3= -4h+2\ii\Vert \vec{e}\Vert\,.
\end{align}
Then we find the solution:
\begin{equation}
    \vec{a}(t) = C_1 \vec{e} \ e^{-4ht} + C_2\vec{v}_2 e^{-4ht -2\ii \Vert \vec{e} \Vert t} + C_3\vec{v}_3 e^{-4ht +2\ii \Vert \vec{e} \Vert t}\,,
\end{equation}
where the coefficients $C_1$, $C_2$, and $C_3$ are to be determined by an initial condition $\vec{a}(0)$.

\subsection{Physical illustrative example}
\begin{figure}
    \centering
    \includegraphics[width=0.49\linewidth]{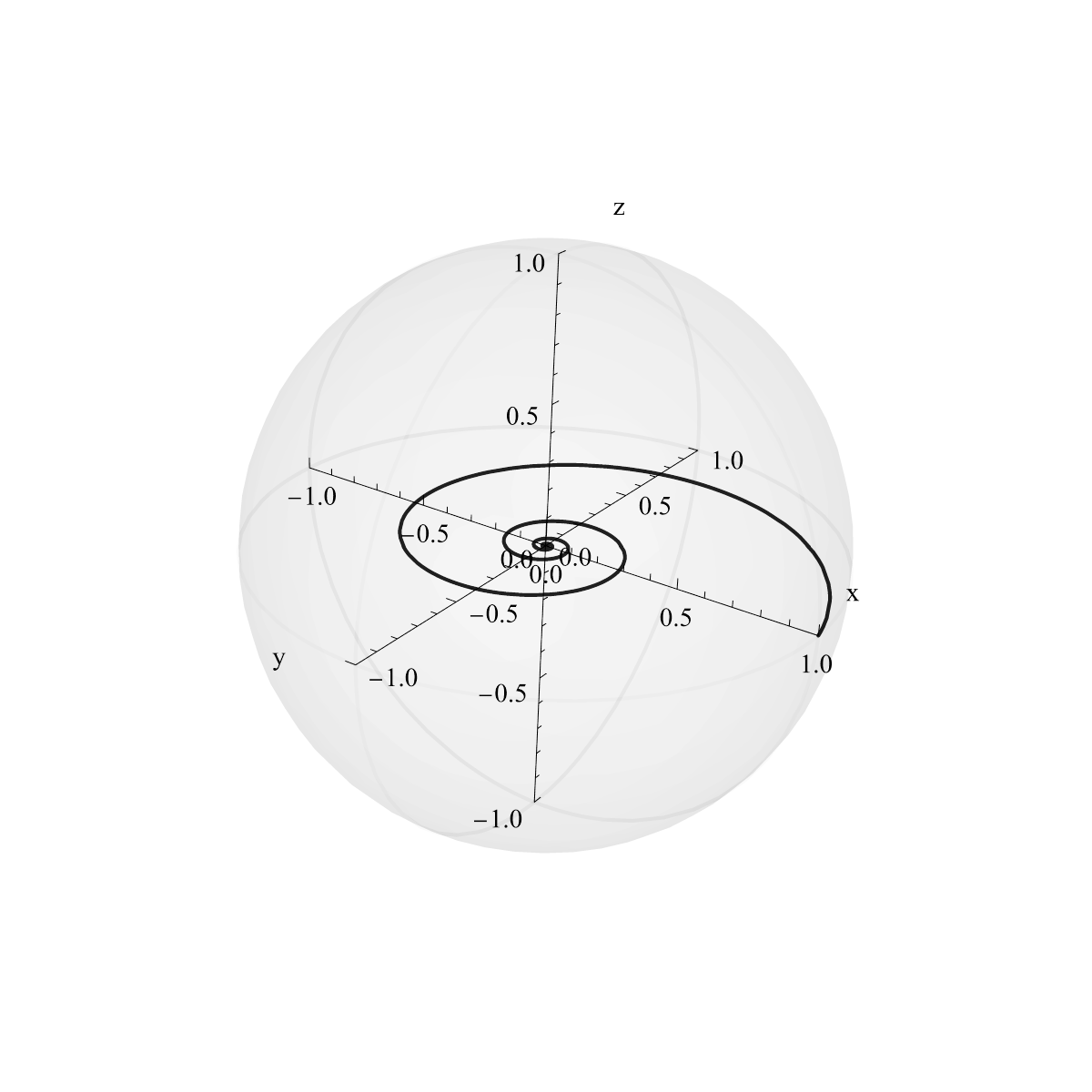}
    \includegraphics[width=0.49\linewidth]{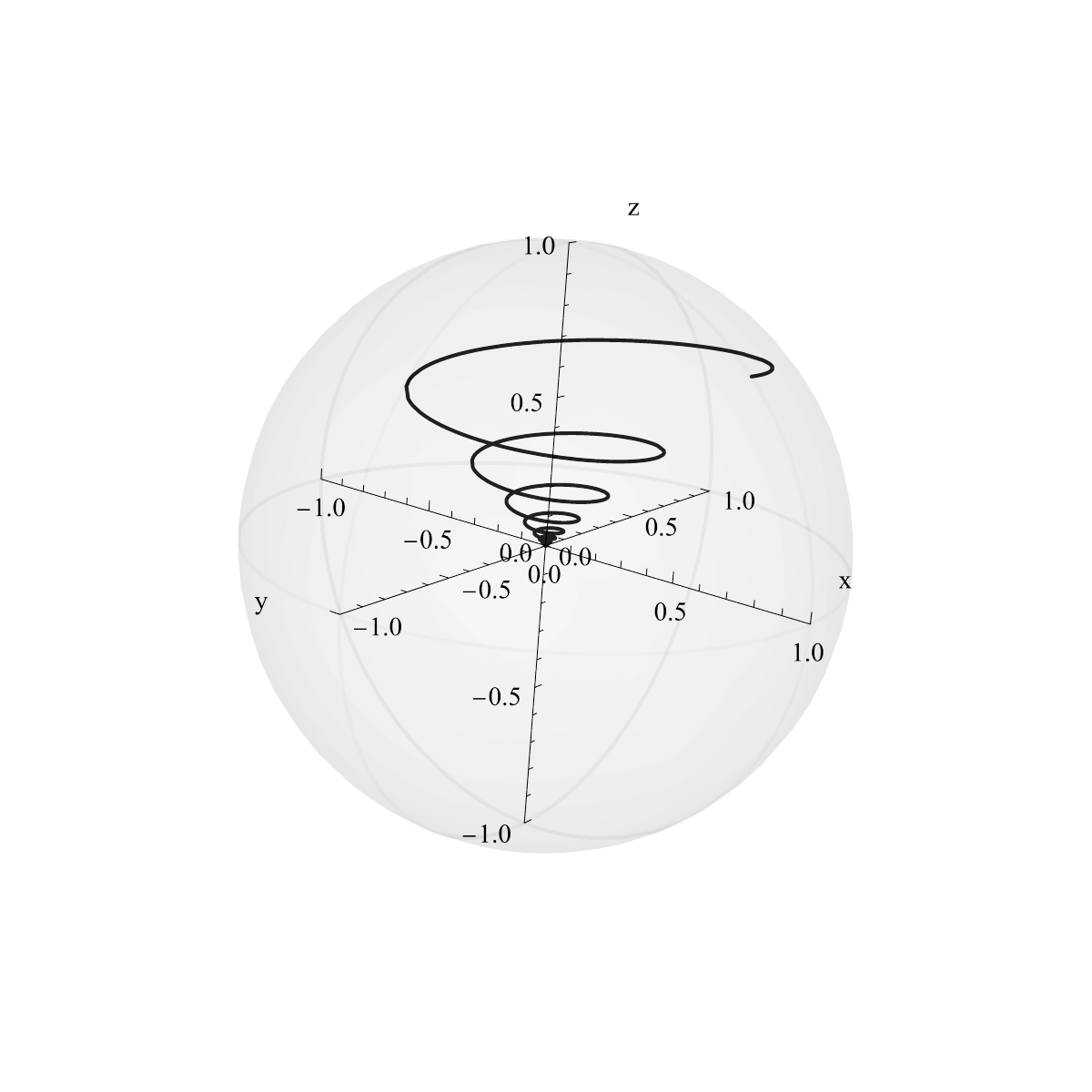}
    \caption{Trajectory of the quantum state given by Equations \eqref{eq: Initial100}\eqref{eq: Initial101} related to the initial condition (left) $\vec{a}(0)=(1,0,0)^T$ and (right) $\vec{a}(0)=(1/\sqrt{2})(1,0,1)^T$ , with the Hamiltonian $H=\omega_0\sigma_z$, where $\omega_0=10$, $h=1$.}
    \label{fig: Initial100}
\end{figure}

In this subsection, we consider a spin-$1/2$ particle interacting with a magnetic field $\vec{B}=\omega_0\hat{z}$.
So, the Hamiltonian is $H=\omega_0\sigma_z$, which gives $\vec{e}=(0,0,\omega_0)^T$. Setting initial condition $\vec{a}(0)=(1,0,0)^T$ we have the solution
\begin{equation}\label{eq: Initial100}
    \vec{a}(t)= e^{-4ht}\left(\cos 2\omega_0 t,\ \sin 2\omega_0 t,\ 0 \right)^T\,,
\end{equation}
and the trajectory can be visualized in Figure \ref{fig: Initial100}. The second example is related to the initial condition $\vec{a}(0)=(1/\sqrt{2})(1,0,1)$, which yields the following solution:
\begin{equation}\label{eq: Initial101}
    \vec{a}(t)= \frac{e^{-4ht}}{\sqrt{2}}\left(\cos 2\omega_0 t,\ \sin 2\omega_0 t,\ 1 \right)^T \, .
\end{equation}
The time evolution of the state vector (Eq. \eqref{eq: Initial101}), as depicted by its trajectory on the Bloch sphere in Fig. \ref{fig: Initial100}, reveals a precessional motion coupled with an inward spiral. This behavior is similar to a damped, rotating system. Consequently, this representative case serves as an example for illustrating the interplay between unitary evolution, which manifests as rotation driven by the Hamiltonian, and non-unitary, dissipative evolution, which is responsible for the observed damping effect.

\begin{figure}
    \centering
    \includegraphics[width=0.99\linewidth]{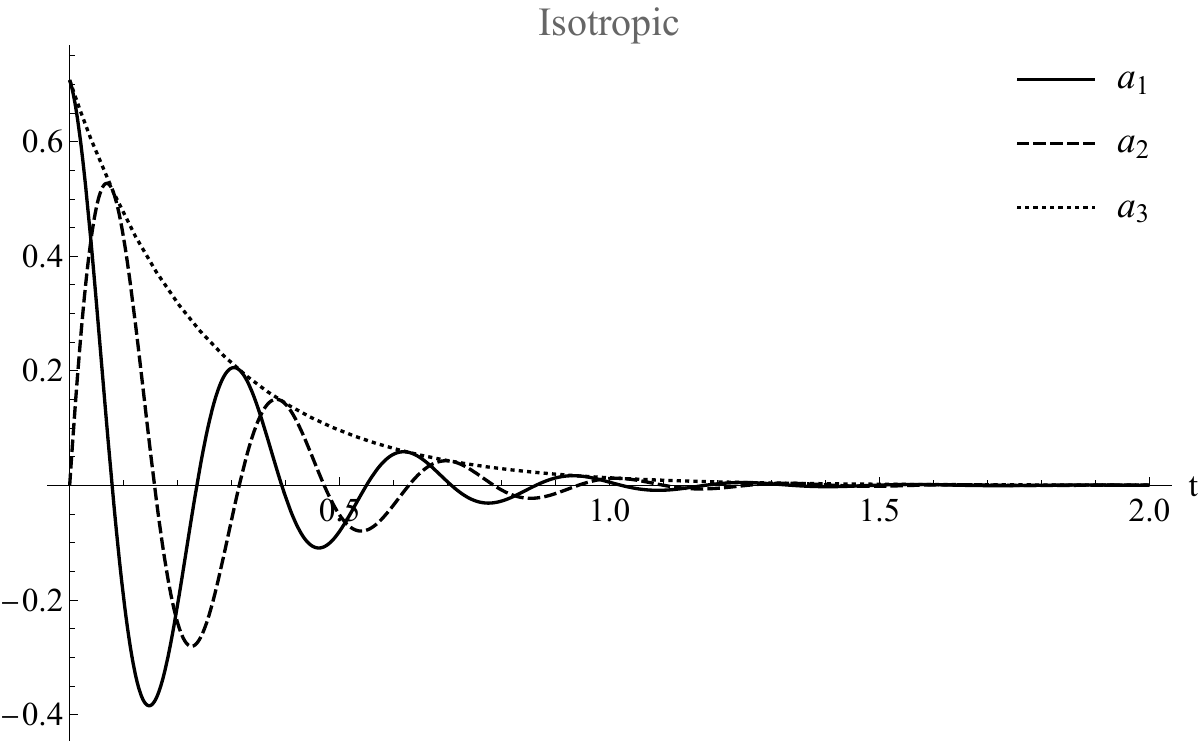}
    \caption{Evolution of the components $a_i(t)$ of the example related to equation \eqref{eq: Initial101}, related to the initial condition $\vec{a}(0)=(1/\sqrt{2})(1,0,1)$, with environment isotropic interactions}
    \label{fig: ComponentsIsotropicInitial101}
\end{figure}

\subsection{Anisotropic case}
\begin{figure}
    \centering
    \includegraphics[width=0.32\linewidth]{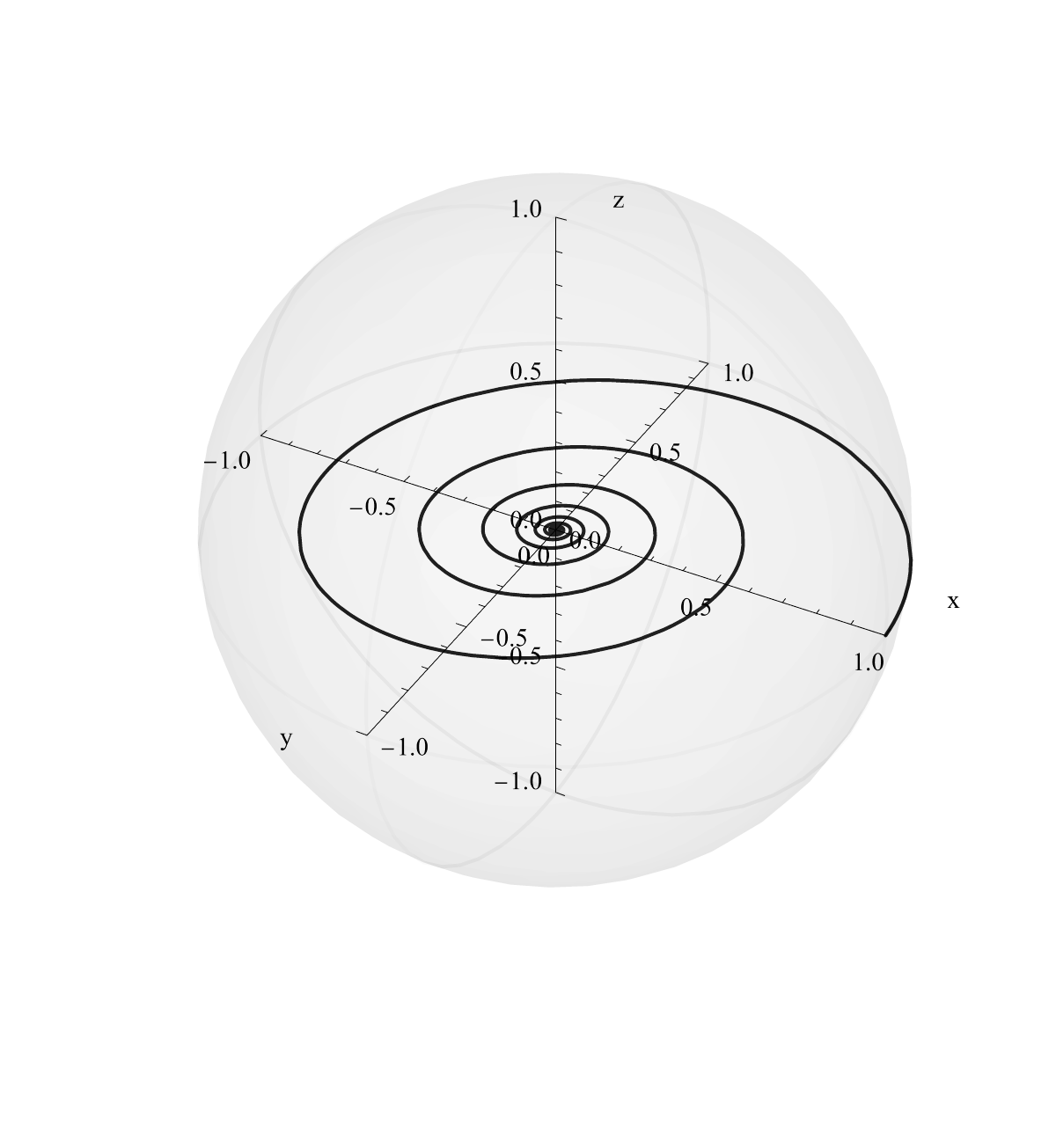}
    \includegraphics[width=0.32\linewidth]{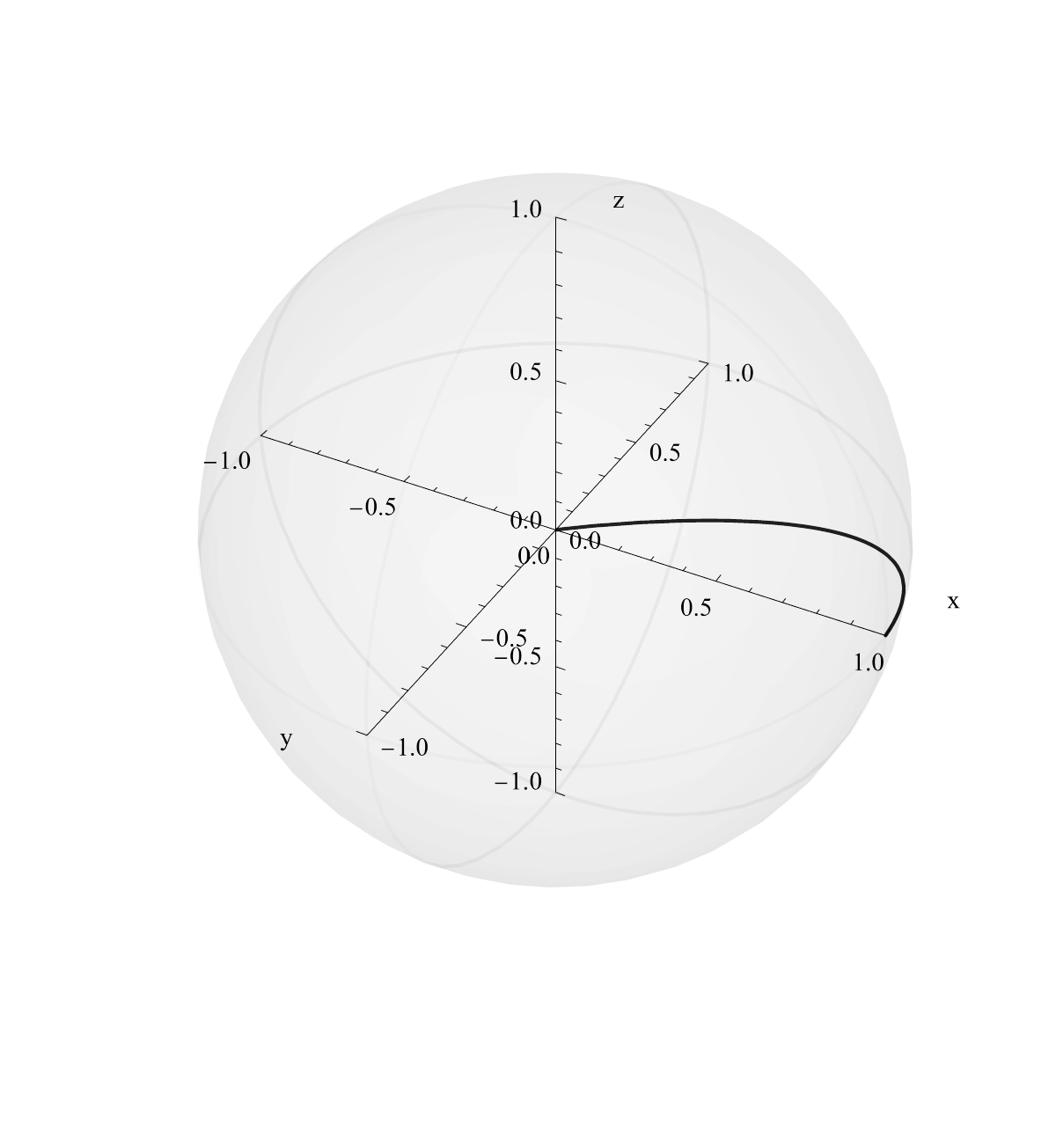}
    \includegraphics[width=0.32\linewidth]{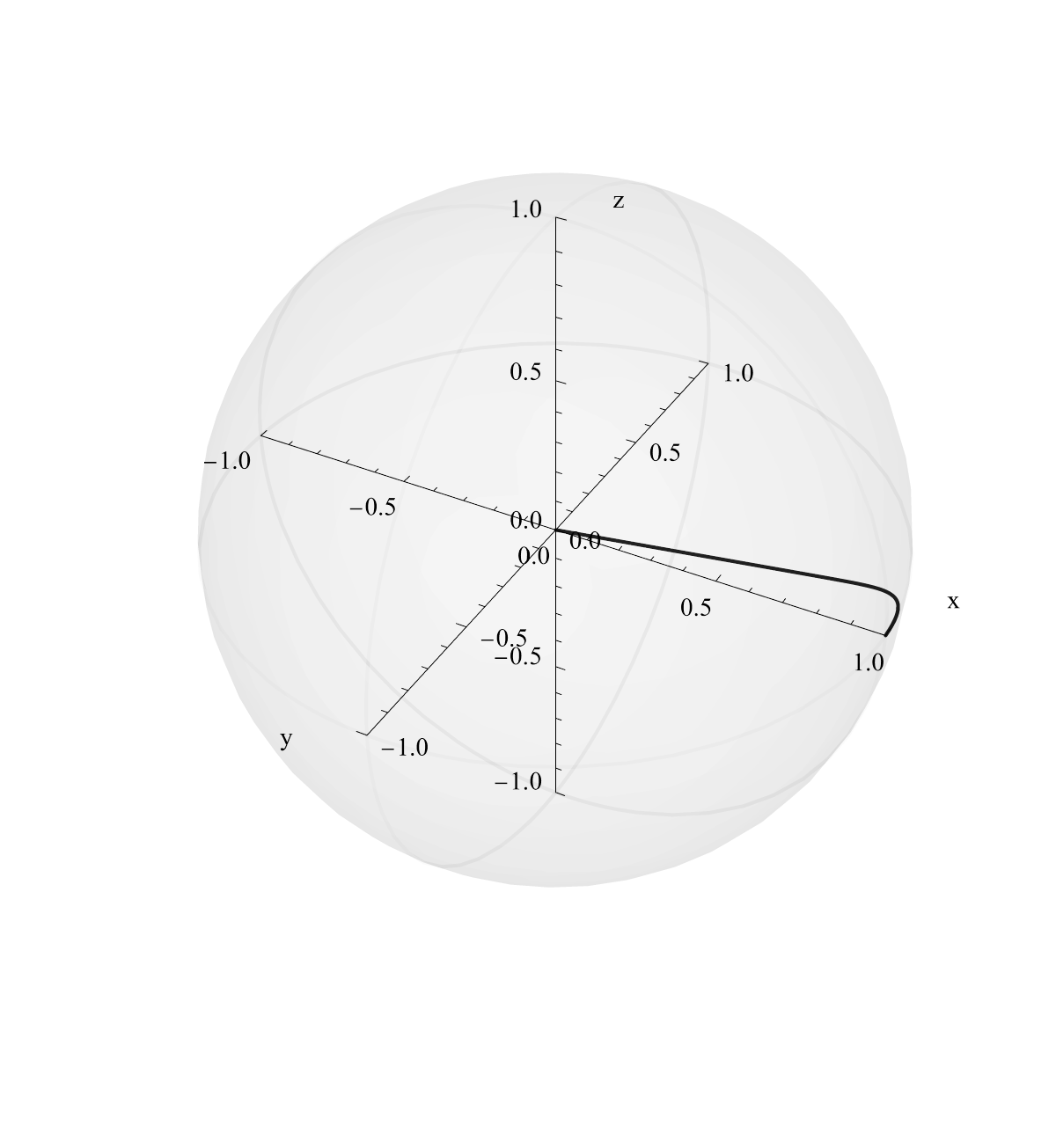}
    \caption{Trajectory of the system in the Bloch ball, for the three regimes (left) oscillation/sub-critical $\beta<1$, (center) critical $\beta=1$, and super-critical $\beta>1$. In the three examples, the contribution of the Hamiltonian is set to $\omega_0=10$, while the decay rates are (left) $h=2$, (center) $h=20$, and (right) $h=50$. The initial condition is the same for the three cases $\vec{a}(0)=(1,0,0)^T$.}
    \label{fig: competition}
\end{figure}

Defining $h_1=h$, and $h_2=h_3=0$, and the Hamiltonian $H=\omega_0 \sigma_z$, gives
\begin{equation}\label{eq: Generator}
    T_d=-2\begin{pmatrix}
        0 & \omega_0 & 0 \\
        -\omega_0 & h & 0 \\
        0 & 0 & h
    \end{pmatrix}\,.
\end{equation}
It is convenient to introduce the parameter $\beta=h/2\omega_0$ and  its pseudo-Lorentz companion $\widetilde{\gamma}= \frac{1}{\sqrt{\beta^2-1}}$.
The eigenvectors and eigenvalues of $T_d$ are then written as 
\begin{align}
    \vec{v}_4 &= (0,0,1)^T\, , &\quad \lambda_4 = -2h \, , \\
    \vec{v}_5 &= (\beta-1/\widetilde{\gamma},\ 1,\ 0)^T\,,  &\quad \lambda_5=2\omega_0(-\beta- 1/\widetilde{\gamma} )\, , \\
    \vec{v}_6 &= (\beta+1/\widetilde{\gamma},\ 1,\ 0)^T\,, &\quad \lambda_6=2\omega_0(-\beta +1/\widetilde{\gamma} ) \, .
\end{align}
One obtains
\begin{equation}
    \vec{a}(t)=C_4\vec{v}_4e^{-2ht} +C_5\vec{v}_5e^{-ht-2\frac{\omega_0}{\widetilde{\gamma}}t}+C_6\vec{v}_6e^{-ht+2\frac{\omega_0}{\widetilde{\gamma}}t}\,. 
\end{equation}
Supposing an initial condition $\vec{a}(0)=(1,\ 0,\ 0)^T$, we get the solution
\begin{equation}\label{eq: soludifferentregimes}
    \vec{a}(t)=\left(e^{-ht}\left[\cosh \left(2\frac{\omega_0}{\widetilde{\gamma}}t\right)+\widetilde{\gamma}\beta \sinh\left(2\frac{\omega_0}{\widetilde{\gamma}}t\right)\right], \ \widetilde{\gamma}e^{-ht}\sinh\left(2\frac{\omega_0}{\widetilde{\gamma}}t\right) ,\ 0\right)^T\,.
\end{equation}
Assuming different values for $\beta$, the system enters different dissipative regimes, such as oscillatory $\beta<1$, damped $\beta>1$, and critical $\beta=1$. These differences can be visualized in Figure \ref{fig: competition}.

The subcritical damping is achieved by choosing $\beta<1$, where the qubit behaves like an oscillatory system with damping. By introducing the Lorentzian companion $\gamma= 1/\sqrt{1-\beta^2}$ the equation \eqref{eq: soludifferentregimes} reads as
\begin{equation}
    \vec{a}(t)=\left(e^{-ht}\left[\cos \left(2\frac{\omega_0}{\gamma}t\right)+\beta \gamma\sin\left(2\frac{\omega_0}{\gamma}t\right)\right], \ \gamma e^{-ht} \sin\left(2\frac{\omega_0}{\gamma}t\right) \,,\ 0\right)^T\,,
\end{equation}
and it is possible to determine the period of the oscillation $\tau=\pi \gamma/\omega_0$.

On the other hand, the super-critical regime is found for the values $\beta>1$. The solution of $\vec{a}(t)$ in equation \eqref{eq: soludifferentregimes} relies on the hyperbolic trigonometric function which can be rewritten as exponentials. Thus, for sufficiently long times, one may factor out the dominant contribution
 $e^{(-h+2\omega_0/\widetilde{\gamma})t}$ which naturally leads to the definition of a characteristic decay time.
 \begin{equation}\label{eq: TimeDecay}
    \tau_d= \frac{(\beta+\sqrt{\beta^2-1})}{2\omega_0}\,.
\end{equation}
For $t\gg\tau_d$ the system behave as $\Vert \vec{a}(t)\Vert \propto e^{-t/\tau_d}$, see the details at appendix \ref{ap: longtimes}. 

The last regime appears at the critical point $\beta=1$, i.e., $h=2\omega_0$, which has the solution
\begin{equation}
    \vec{a}(t)=e^{-2\omega_0 t}\left(1+2\omega_0 t, \ 2\omega_0 t, \ 0 \right)^T\,.
\end{equation}

\begin{figure}[]
    \centering
    \includegraphics[width=.99\linewidth]{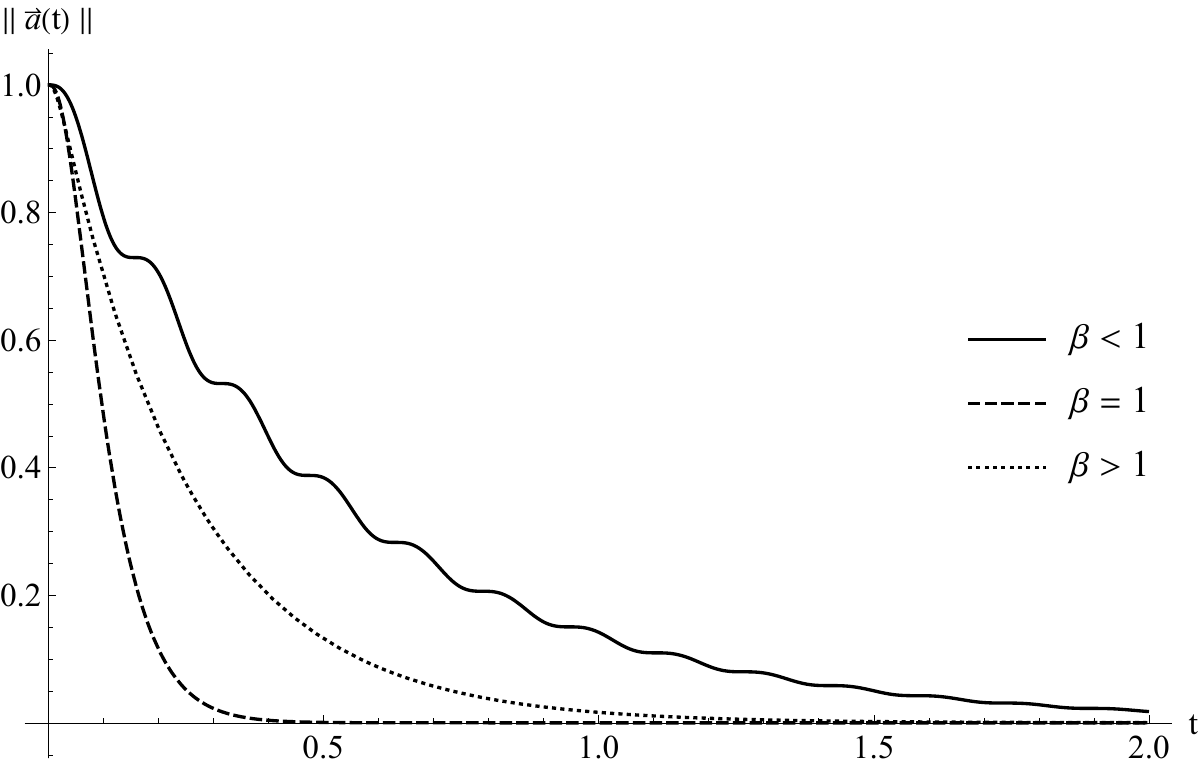}
    \caption{Decay comparison of the examples in Fig. \ref{fig: competition},  for the three regimes oscillation/sub-critical $\beta<1$, critical $\beta=1$, and super-critical $\beta>1$. The system related to the critical case decays faster then the others.}
    \label{fig: decaycomparison}
\end{figure}

\begin{figure}
    \centering
    \includegraphics[width=0.55\linewidth]{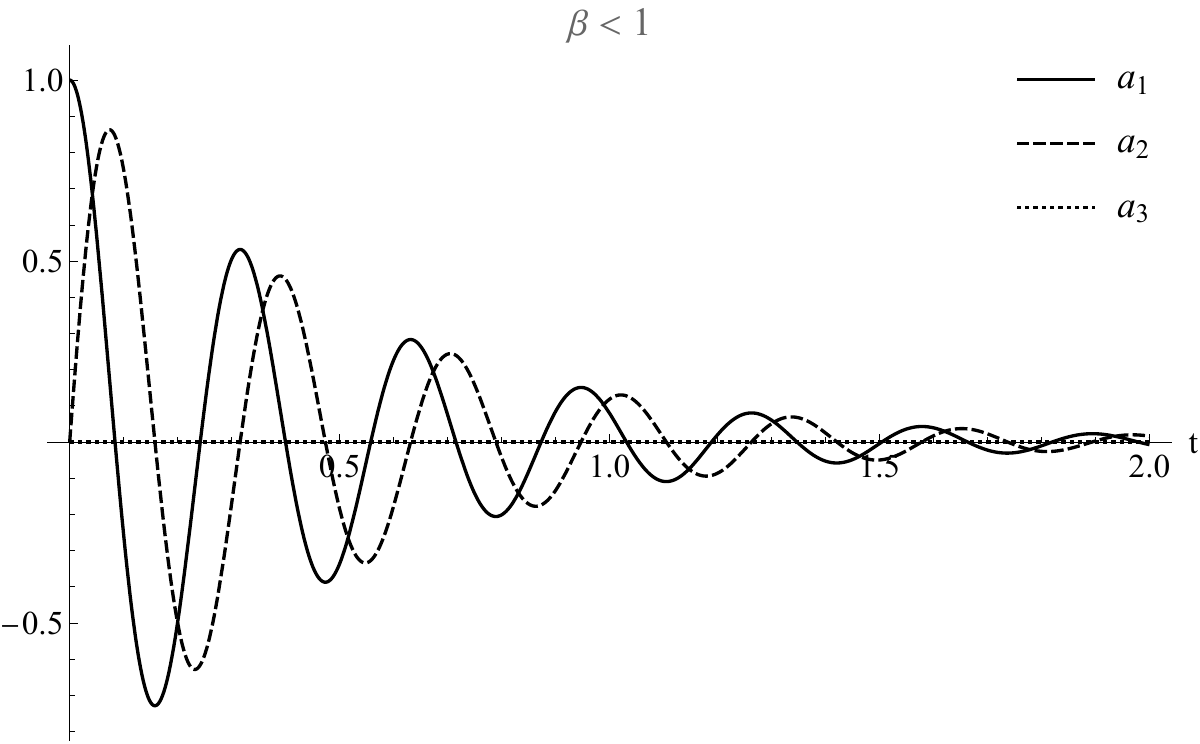}
    \includegraphics[width=0.55\linewidth]{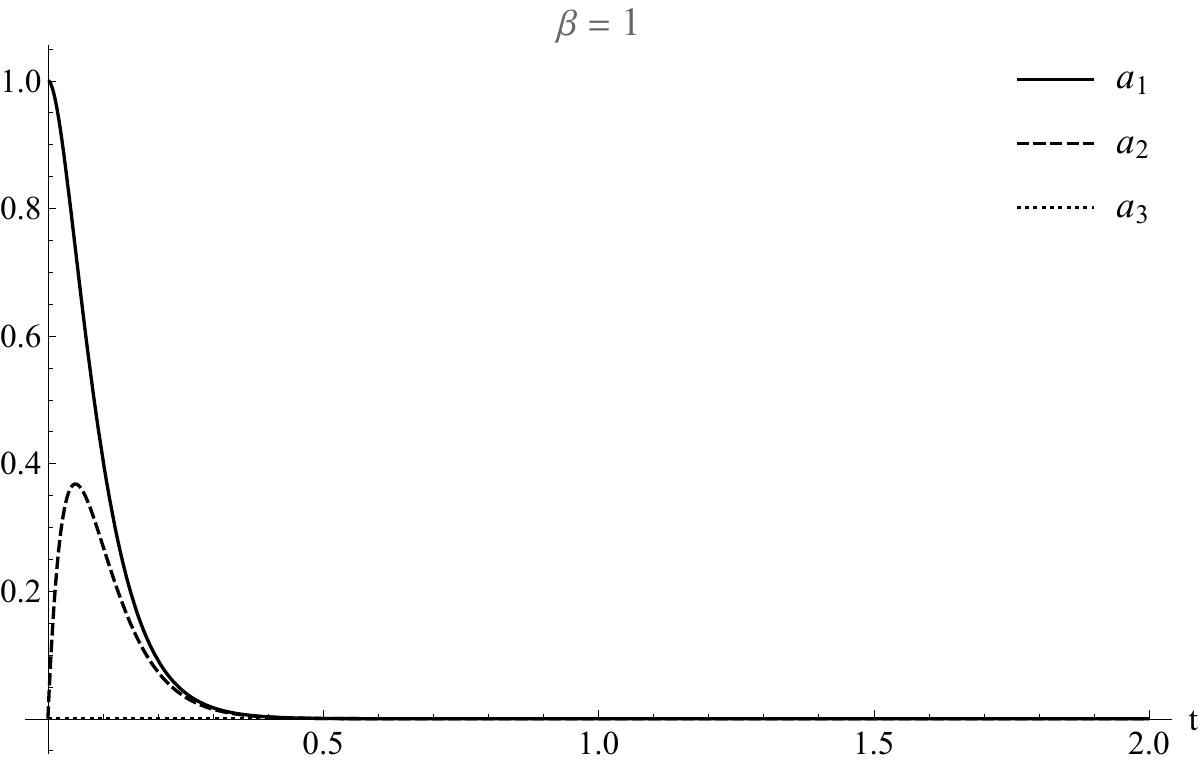}
    \includegraphics[width=0.55\linewidth]{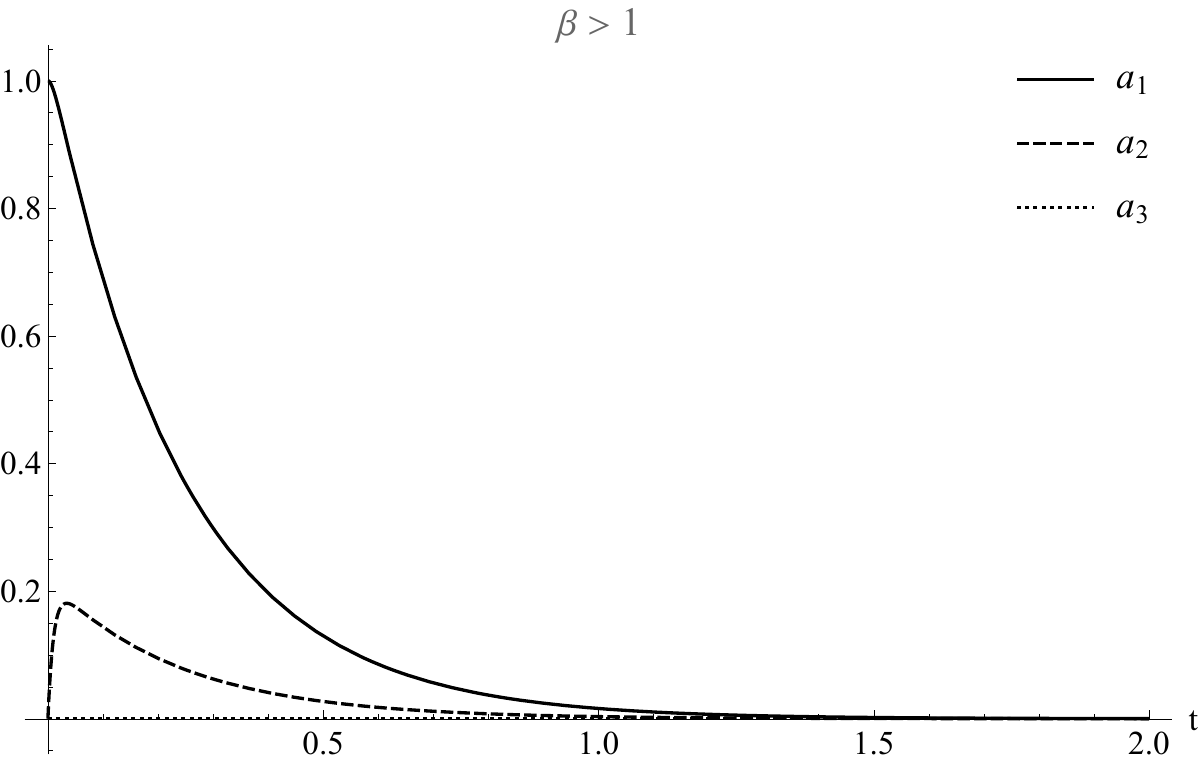}
    \caption{Plot of the evolution of the components $a_i(t)$ of the $\vec{a}(t)$ associated to the Pauli matrices $\sigma_i$ with $i=1,2,3$,  for the (top) subcritical, (center) critical, and (bottom) supercritical regimes. }
    \label{fig: componentsEVO}
\end{figure}

This system exhibits a behaviour closely analogous to that of a classical damped harmonic oscillator, in which the most rapid decay occurs in the critically damped regime. This feature is clearly visible in Fig. \ref{fig: decaycomparison} (and for each Bloch component in Fig. \ref{fig: componentsEVO}). 
The analogy stems from the competition between the unitary Hamiltonian dynamics and the dissipative influence of the environment: the former induces rotations generated by the interaction along the $z$-axis through the action of $\sigma_3=\sigma_z$, whereas the latter produces 
damping along the $x$-axis via the action of $\sigma_1=\sigma_x$.
For completeness, Fig. \ref{fig: componentsEVO} displays the time evolution of the individual Bloch components for the anisotropic cases, thereby illustrating the different decay regimes.

\subsection{One direction} \label{subsec: onedirection}
Here we explore the behavior of the system when both the Hamiltonian and decay rate are related to $\sigma_z$. So, $H=\omega_0 \sigma_z$, $h_1=h_2=0$, while $h_3=h$, and
\begin{equation}\label{eq: Generatorrotz}
    T_d=-2\begin{pmatrix}
        h & \omega_0 & 0 \\
        -\omega_0 & h & 0 \\
        0 & 0 & 0
    \end{pmatrix}\,.
\end{equation}
Therefore, the action of the matrix $T_d$ over the Bloch vector $\vec{a}$ only affects the $\sigma_x$ and $\sigma_y$ directions, leaving the $\sigma_z$ unchanged. Along these lines, a state that is initially $\vec{a}(0)=(0,0,1)^T$ will not change over time, which can be represented as a fixed point of the GKSL equation. 

\begin{figure}
    \centering
    \includegraphics[width=0.5\linewidth]{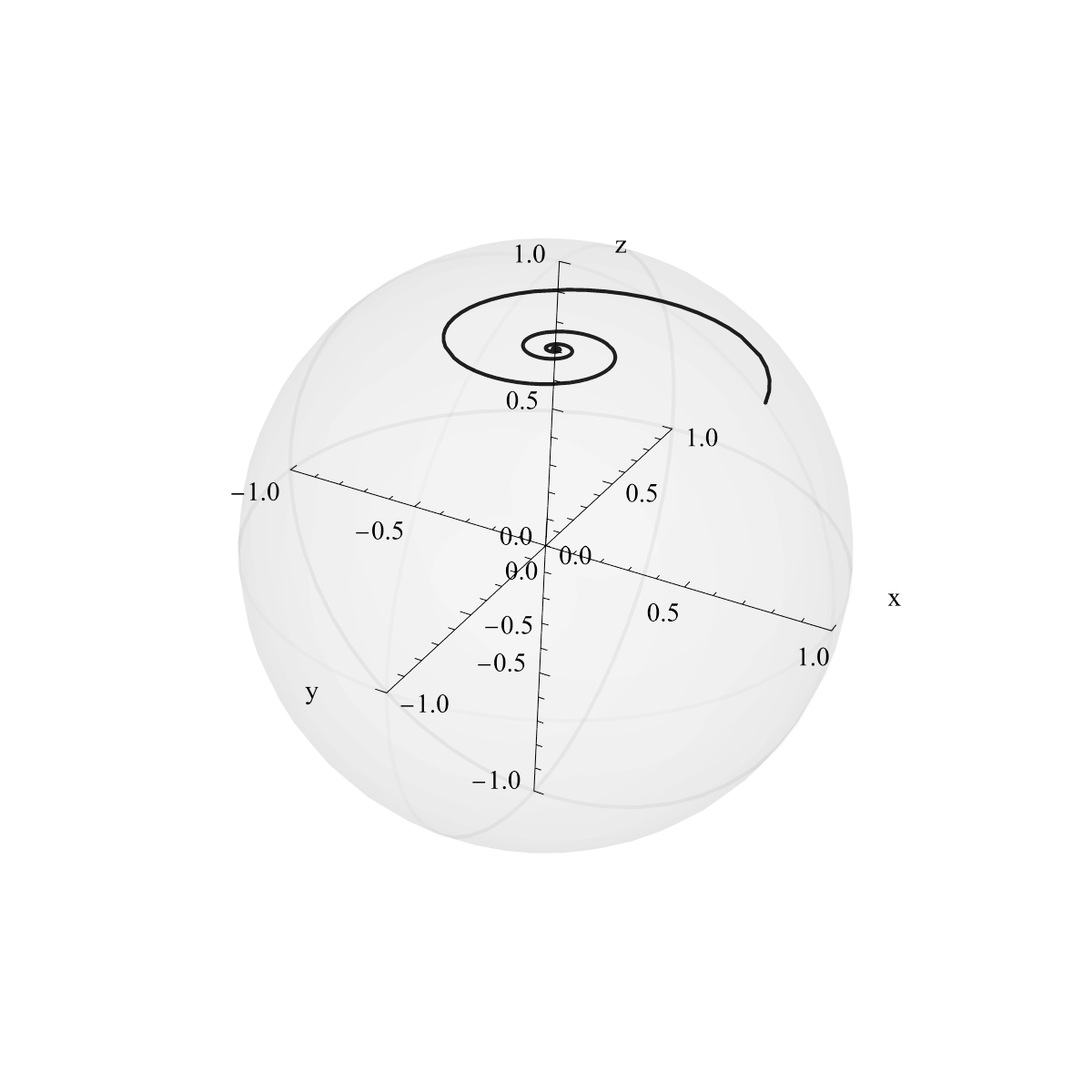}
    \caption{Trajectory of the state given by Eq. \eqref{eq: decayz} where $\omega_0=5$ and $h=1$. The initial condition is $\vec{a}(0)=(1/\sqrt{2})(1,0,1)^T$. In this example, the $\sigma_z$ direction does not change over time.}
    \label{fig: decayz}
\end{figure}

On the other hand, a state that has $\sigma_x$ or $\sigma_y$ components will rotate and decay, for example the state $\vec{a}(0)=(1,0,1)^T/\sqrt{2}$ will evolve according to 
\begin{equation}\label{eq: decayz}
    \vec{a}(t)=\frac{1}{\sqrt{2}}\left(e^{-2ht}\cos2\omega_0t \ ,\ e^{-2ht}\sin2\omega_0t \ , \ 1\right)^T \, ,
\end{equation}
and it can be visualized in Fig. \ref{fig: decayz}, while the $\sigma_z$ component remains constant.

This example shows that a state with components only along the $\sigma_z$ direction is a fixed point of the evolution, and it can be seen in the solution before setting the initial condition
\begin{equation}
    \vec{a}(t)=\left( -\ii c_2 e^{-2t(h+i\omega_0) }+\ii c_3 e^{-2t(h-i\omega_0) }\ ,\ c_2 e^{-2t(h+i\omega_0) }+ c_3 e^{-2t(h-i\omega_0) } \ , \ c_1\right)^T \, ,
\end{equation}
and setting $c_2=c_3=0.$
More generally, each point in the z-axis acts as an attractor for the dynamics in xy-plane. In other words, any initial state will decohere in the xy-plane while preserving its z-component, eventually converging to a final state on the z-axis. This line of fixed points represents the set of stable equilibrium states for the qubit under this specific Hamiltonian and dissipative channel.
This phenomenon illustrates a strategy for enhancing qubit stability: by engineering the dominant environmental interaction to create a (almost) decoherence-free subspace, the computational basis states (i.e, $|\uparrow\rangle$ and $|\downarrow\rangle$ as eigenstates of $\sigma_z$) become robust  states, that act like a pointer state.

\subsection{Exceptional points: discussion} 

In a broader context, the exceptional points (EPs) \cite{EP, Miri2019} are non-Hermitian degeneracies at which both eigenvalues and their corresponding eigenvectors coalesce, leading to a breakdown of the conventional Hermitian spectral theory. In quantum systems, EPs emerge when dissipation or gain introduces non-conservativity, rendering the effective Hamiltonian non-Hermitian. This results in a characteristic spectral topology, where eigenvalues exhibit square-root branch point behavior. For a single qubit subject to engineered loss, such as in superconducting circuits or optical platforms, the Hamiltonian can be tailored to realize an EP by balancing coherent coupling and decay rates. In this regime, the qubit exhibits a non-trivial evolution marked by spontaneous symmetry breaking, mode non-orthogonality, and enhanced sensitivity to perturbations. Importantly, recent experimental realizations have demonstrated quantum state tomography across EPs \cite{Naghiloo2019}, revealing distinctive features such as the coalescence of decay rates at the EP, the transition between underdamped and overdamped dynamics, and signatures of PT-symmetry breaking in the relaxation pathways of the qubit. 

In the anisotropic single-qubit model presented in this section, the effective Bloch evolution is generated by the non-Hermitian matrix $T_d$, Eq. \eqref{eq: Generator}. The eigenvalues of $T_d$ determine three dynamical regimes (sub-critical, critical, and super-critical), analogous to a damped harmonic oscillator. At the critical point ($\beta=1$), two eigenvalues and eigenvectors coalesce, indicating a Liouvillian exceptional point. In another words, a non-Hermitian degeneracy  occurs in the Liouvillian superoperator of the open quantum system. The dynamics near this point are qualitatively distinct, with the fastest decay consistent with known non-Hermitian degeneracies.

\section{ SU$(2)$ symmetry content of GLKS dynamics}\label{SU2open}
In this section we adopt an approach to the GLKS system based on the underlying SU$(2)$ symmetry and related tools. Although we (artificially) leave linearity of the system, we unveil a useful separation between the angular regime, which involves the whole Lindblad operator  $\mathcal{L}$, and  the radial regime rule, which is explicitly ruled by $\mathcal{L}_{\mathrm{diss}}$ only. More precisely, the SU$(2)$ symmetry approach, actually the left coset $\mathrm{SU}(2)/\mathrm{U}(1)$ symmetry, decouples the  dynamics of the system. It provides a link between the change in the state's purity, the decay rates, and the evolution of its orientation and on the Bloch sphere as a nonlinear dynamical system.

\subsection{SU$(2)$ content of density matrix}

An arbitrary  density operator in $\C^2$ has the following spectral decomposition
\begin{equation}
\label{densspectr}
\rho= p_1 \mathrm{P}_{\bu_1} + p_2 \mathrm{P}_{\bu_2}\, ,  \quad p_1+p_2=1\,  ,
\end{equation}
where the $p_i \in [0,1]$ have a classical probability meaning and the $\mathrm{P}_{\bu_i}=|\bu_i\rg\lg\bu_i|$ are orthogonal projector (pure states) on  eigendirections $\bu_i$, $i=1,2$. 
We suppose that there is no degeneracy and $p_1 > p_2 >0$ (generic situation). By introducing the element of $U(\theta, \phi)\in \mathrm{SU}(2)/\mathrm{U}(1)$ which diagonalizes $\rho$, the later can be written as 
\begin{equation}
\label{eq: densopersu2}
\rho\equiv \rho_{r,\theta,\phi} = \frac{1}{2}\left(\bun_2 + r U(\theta, \phi)\sigma_3U^{\dag}(\theta, \phi)\right) \equiv \frac{1}{2}\left(\bun_2 + r \sigma_{\theta,\phi}\right)\,, 
\end{equation}
with
\begin{equation}
\label{Umat}
U(\theta, \phi):= \begin{pmatrix}
  \cos\frac{\theta}{2}    &  -  \sin\frac{\theta}{2} \,e^{-\ii\phi} \\
   \sin\frac{\theta}{2} \,e^{\ii\phi}   &   \cos\frac{\theta}{2}
\end{pmatrix}\, , \quad \theta\in [0,\pi]\,,\quad  \phi\in [0,2\pi)\, , 
\end{equation}
 and $r= 2p_1-1\in [0,1]$ is the normalized excess of $p_1$ above the uniform baseline. 
 The matrix $\sigma_{\theta,\phi}$ in \eqref{eq: densopersu2} is a traceless involution  (involution means its square is the identity)  in SU$(2)$:
\begin{equation}
\label{sigtheph}
\sigma_{\theta,\phi}= U(\theta,\phi) \sigma_3U^{\dag}(\theta,\phi)= \begin{pmatrix}
   \cos\theta   &  \sin\theta \,e^{-\ii\phi}  \\
  \sin\theta\, e^{\ii\phi}    &  -\cos\theta
\end{pmatrix}\, , \quad \sigma_{\theta,\phi}= \sigma^{\dag}_{\theta,\phi}\, ,\quad \sigma^2_{\theta,\phi}=\bun_2\,. 
\end{equation}
We check that $(\theta,\phi)\mapsto  U(\theta, \phi)$ is a representation of the half-circle, \begin{equation}
\label{Urep}
U(\theta, \phi)U(\theta^{\prime}, \phi)= U(\theta + \theta^{\prime}, \phi)\, , 
\end{equation}
This property is consistent with the fact  that $U(\theta, \phi)$ is an element of the left coset $\mathrm{SU}(2)/\mathrm{U}(1)$, since any right U$(1)$ factor $\begin{pmatrix}
   e^{\ii\omega}   &  0  \\
   0   &  e^{-\ii\omega}
\end{pmatrix}$ of $U(\theta, \phi)$ in \eqref{eq: densopersu2} is transparent due to the fact it commutes with $\sigma_3$. 
Its column vectors 
\begin{equation}
\label{bu12}
\bu_1= \begin{pmatrix}
       \cos\frac{\theta}{2}   \\
        \sin\frac{\theta}{2}e^{\ii\phi}
\end{pmatrix}\, , \quad  \bu_2= \begin{pmatrix}
    -\sin\frac{\theta}{2}e^{-\ii\phi}     \\
      \cos\frac{\theta}{2}   
\end{pmatrix}\, , 
\end{equation}
which form  an orthonormal basis for $\C^2$, are precisely the normalised eigenvectors of $\rho$, consistently with the notation in \eqref{densspectr}. Moreover, they can be viewed as  spin  one-half  coherent states $|\theta,\phi \rg$, defined in terms of spherical coordinates $(\theta,\phi)$ as the quantum counterpart of the classical state $ \hat{\mathbf{n}}(\theta,\phi)$ in the (Bloch) sphere $\mathbb{S}^2$  by
\begin{equation}
	\begin{split}
&\bu_1= \begin{pmatrix}
       \cos\frac{\theta}{2}   \\
        \sin\frac{\theta}{2}e^{\ii\phi}
\end{pmatrix}\equiv |\theta,\phi \rg=\left(\cos\frac{\theta}{2}\,\upa + e^{\ii \phi}\sin\frac{\theta}{2}\,\dwa\right)\\ 
	&=  \begin{pmatrix}
		\cos\frac{\theta}{2}   &- \sin\frac{\theta}{2} e^{-\ii \phi}  \\
		\sin\frac{\theta}{2} e^{\ii \phi}   &  \cos\frac{\theta}{2}
	\end{pmatrix}\begin{pmatrix}
		1   \\
		0 
	\end{pmatrix}
	=U(\theta,\phi)\upa\equiv D^{\frac{1}{2}}\left(\xi^{-1}_{\hat{\mathbf{n}}}\right)\upa \,. 
	\end{split}
	 \label{operatorD} 
\end{equation}
Here, $\xi_{\hat{\mathbf{n}}}$  corresponds, through the homomorphism SO$(3)$ $\mapsto$ SU$(2)$,  to  the specific rotation $\mathcal{R}_{\hat{\mathbf{n}}}$, in the Bloch sphere, which maps the unit vector pointing to the north pole, $\hat{\mbox{\textbf{\textit{k}}}}=(0,0,1)$, to $\hat{\mathbf{n}}$. The operator $D^{\frac{1}{2}}\left(\xi^{-1}_{\hat{\mathbf{n}}}\right)$ represents the element  $\xi^{-1}_{\hat{\mathbf{n}}}$ of SU$(2)$ in its complex two-dimensional unitary irreducible representation. Thus, the spectral decomposition \eqref{densspectr}of $\rho_{r,\theta,\phi}$  is expressed  in terms of the two orthogonal spin CS:
\begin{equation}
\label{densspectr2}
\rho_{r,\theta,\phi}= \frac{1+r}{2} |\theta,\phi \rg\lg\theta,\phi | + \frac{1-r}{2} |\theta +\pi,\phi \rg \lg\theta +\pi,\phi | \,. 
\end{equation}
 We remind here some fundamental properties of these CS:
\begin{equation}
\label{ }
\lg\theta,\phi|\theta,\phi \rg = 1\, , \quad \int_{\mathbb{S}^2}\frac{\ud \bu}{4\pi}\,|\theta,\phi \rg\lg \theta,\phi \rg = \bun_2\, , \quad \ud \bu = \sin\theta\, \ud \theta\,\ud \phi\, .
\end{equation}
Note the extreme cases, $r= 1$ (pure state $\equiv$ spin coherent state), and $r=0$ (totally random state):
\begin{equation}
\label{r01}
\rho_{1,\theta,\phi}= \left\vert \theta,\phi\right\rangle \left\langle \theta,\phi\right\vert\, , \quad \rho_0= \frac{1}{2}\bun_2\,. 
\end{equation}

\subsection{GLKS system in terms of  parameters $(r,\theta,\phi)$}
\subsubsection{Radial dissipation}
We now formulate the GLKS system \eqref{eq: lindbladA} in functions of the  Bloch ball parameters $(r,\theta,\phi)$. 
First, by using the expression \eqref{eq: densopersu2}, we have for the time derivative of  $\rho_{r,\theta,\phi}$ 
\begin{equation}
\label{rhodot}
\frac{\ud \rho_{r,\theta,\phi}}{\ud t}= \frac{\dot r}{2}\, \sigma_{\theta,\phi} + \frac{r}{2}\, \frac{\ud \sigma_{\theta,\phi}}{\ud t}\, .
\end{equation}
Now, from \eqref{sigtheph} and the identities $UU^{\dag}=\bun_2= U^{\dag}U \Rightarrow \dot{U}U^{\dag}=-U\dot{U}^{\dag}$, we get
\begin{equation}
\label{ }
\frac{\ud \sigma_{\theta,\phi}}{\ud t}= \dot{U}\sigma_3U^{\dag} + U\sigma_3 \dot{U}^{\dag}= \left[\dot{U}U^{\dag},\sigma_{\theta,\phi}\right]\, . 
\end{equation}
Thus the  Lindblad equation takes the form
\begin{equation}
\label{GLKS2}
 \frac{\dot r}{2}\, \sigma_{\theta,\phi} +  \frac{r}{2}\,\left[\dot{U}U^{\dag},\sigma_{\theta,\phi}\right]= -\ii \frac{r}{2}\,\left[H,\sigma_{\theta,\phi}\right] + \frac{r}{2}\sum_{k=1}^3 h_k\left(\sigma_k\,\sigma_{\theta,\phi}\,\sigma_k - \sigma_{\theta,\phi}\right)\,. 
\end{equation}
By using $\sigma^2_{\theta,\phi}= \bun_2$, 
\begin{equation}
\label{rdotr}
\frac{\dot{r}}{r}\bun_2 = \left[\sigma_{\theta,\phi},\dot{U}U^{\dag} +\ii H\right]\sigma_{\theta,\phi} + \sum_{k=1}^3 h_k\left[\left(\sigma_k\sigma_{\theta,\phi}\right)^2- \bun_2\right]\,. 
\end{equation}
Taking the trace of this equation and using $\mathrm{Tr}\left(\left[\sigma_{\theta,\phi},A\right]\sigma_{\theta,\phi}\right)=\mathrm{Tr}\left(\sigma_{\theta,\phi}A\sigma_{\theta,\phi} - A\right)=0$ for all $A\in \mathrm{M}(2,\R)$, we get the equation: 
\begin{equation}
\label{rdr}
\begin{split}
\frac{\dot{r}}{r} &= \frac{1}{2} \left(\sum_{k=1}^3 h_k\left(\mathrm{Tr}\left(\sigma_k\sigma_{\theta,\phi}\right)^2 -2\right)\right)\\
&=-2\left[ h_1(1-\sin^2\theta\cos^2\phi)+h_2(1-\sin^2\theta\sin^2\phi) + h_3\sin^2\theta  \right]\,. 
\end{split}
\end{equation}
We note that there is no explicit dependence of the Hamiltonian. The action of the latter is implicit only through the time dependence of the angles $\theta$ and $\phi$. 
Since the decay rates $h_k$  are non-negative, the entire expression on the right-hand side is always less than or equal to zero.
Henceforth, Eq. \eqref{rdr} is a fundamental result that directly connects the change in the state's purity to the decay rates. The parameter $r$ is a measure of the state's purity, where $r=1$ is a pure state and $r=0$ is a maximally mixed state. This proves that  $\dot r \leq 0$, meaning the purity of the state can only decrease or remain constant over time. The fact that $\dot r \leq 0$ confirms the principle of dissipation, as the system loses coherence to its environment. 
This interpretation  is confirmed with the increase of the von Neumann entropy: 
\begin{equation}
\label{VNentr}
S_{\rho}:= -\mathrm{Tr}(\rho\,\ln\rho)= -\frac{1+r}{2} \ln\frac{1+r}{2} - \frac{1-r}{2} \ln\frac{1-r}{2}\, , \quad  \dot{S_{\rho}}= -\frac{\dot{r}}{2} \ln\frac{1+r}{1-r}\geq 0\,. 
\end{equation}

\subsubsection{Angular dynamical system}

Let us now decouple the angular variables motions from the radial one, starting  from \eqref{rdotr} and using \eqref{rdr}. By introducing the two traceless matrix expressions:
\begin{align}
\label{MN}
M(\theta,\phi)&=  \sum_{k=1}^3 h_k\left[\left(\sigma_k\sigma_{\theta,\phi}\right)^2 -\frac{1}{2}\left(\mathrm{Tr}\left(\sigma_k\sigma_{\theta,\phi}\right)^2 \right)\bun_2\right]\, , \quad \mathrm{Tr} \, M(\theta,\phi)= 0\, , \\
N(\theta,\phi)&= -\ii \left[H,\sigma_{\theta,\phi}\right]\,\sigma_{\theta,\phi} + M(\theta, \phi)\,, 
\end{align} 
we then obtain the matrix dynamical system for angular variables $(\theta,\phi)$:
\begin{equation}
\label{dotUsig}
\left[\dot{U}U^{\dag}, \sigma_{\theta,\phi}\right]\sigma_{\theta,\phi} = N(\theta,\phi)\,. 
\end{equation} 
We now make explicit $\dot{U}$ and $\dot{U}U^{\dag}\equiv \dot X$ in terms of $\dot\phi$ and $\dot\theta$ from the matrix \eqref{Umat}. Here $X$ is  the antihermitian ``logarithm''  of $U$:
\begin{equation}
\label{eX}
U=e^{X}\, , \quad X\in \mathfrak{su}(2)\, ,\quad \dot{U}U^{\dag} = \dot X\, . 
\end{equation}
\begin{equation}
\label{dotU}
\begin{split}
\dot{U}&= \ii \dot\phi\, \sin\frac{\theta}{2}\,\begin{pmatrix}
  0    &  e^{-\ii\phi}  \\
   e^{\ii\phi}   &  0
\end{pmatrix} + \frac{\dot\theta}{2}\begin{pmatrix}
 -\sin\frac{\theta}{2}     & - e^{-\ii\phi}\cos\frac{\theta}{2}  \\
 e^{\ii\phi}\cos\frac{\theta}{2}    &  -\sin\frac{\theta}{2} 
\end{pmatrix}\\
&= \ii \dot\phi\,\sin\frac{\theta}{2}\,\mathrm{n}(\phi) + \frac{\dot\theta}{2}\,U(\theta+\pi,\phi)\, , 
\end{split}
\end{equation}
where we have introduced the Hermitian involution
\begin{equation}
\label{nphi}
\mathrm{n}(\phi)= \begin{pmatrix}
 0     &  e^{-\ii \phi}  \\
 e^{\ii \phi}     &  0
\end{pmatrix}\, , \quad  \mathrm{n}^{\dag}(\phi)= \mathrm{n}(\phi)\, , \quad \mathrm{n}^{2}(\phi)= \bun_2\,. 
\end{equation}
Hence we get for \eqref{dotU}:
\begin{equation}
\label{dotX}
\begin{split}
\dot{U}U^{\dag}&=\dot{X}= \ii \dot\phi \, \sin\frac{\theta}{2}\, \begin{pmatrix}
 -  \sin\frac{\theta}{2}    & \cos\frac{\theta}{2}  e^{-\ii\phi}  \\
  \cos\frac{\theta}{2}  e^{\ii\phi}   &   \sin\frac{\theta}{2} 
\end{pmatrix} + \frac{\dot\theta}{2}\,\begin{pmatrix}
 0     & - e^{-\ii\phi}  \\
   e^{\ii\phi}    &  0 
\end{pmatrix}\\ &= \ii \dot\phi\,\sin\frac{\theta}{2}\,\mathrm{F}(\theta,\phi) + \frac{\dot\theta}{2}\,\mathrm{s}(\phi)\, , 
\end{split}
\end{equation}
where we have introduced  the traceless involution matrix
\begin{equation}
\label{Ftht}
\mathrm{F}(\theta,\phi)= \begin{pmatrix}
 -  \sin\frac{\theta}{2}    & \cos\frac{\theta}{2}  e^{-\ii\phi}  \\
  \cos\frac{\theta}{2}  e^{\ii\phi}   &   \sin\frac{\theta}{2} 
\end{pmatrix}\, , \quad \mathrm{F}^2(\theta,\phi)=\bun_2\, , \quad \mathrm{det}\,\mathrm{F}(\theta, \phi)=-1\,, 
\end{equation}
and the traceless anti-involution Hermitian matrix
\begin{equation}
\label{sphi}
\mathrm{s}(\phi) = \begin{pmatrix}
0    & - e^{-\ii\phi}  \\
 e^{\ii\phi}   & 0
\end{pmatrix}\, , \quad  \mathrm{s}^2= -\bun_2\, , \quad \mathrm{det}\,\mathrm{s}(\phi)=1\,. 
\end{equation}
Note the relations preserving the traceless property:
\begin{equation}
\label{sFs}
\mathrm{s}(\phi)\, \mathrm{F}(\theta,\phi)= \begin{pmatrix}
 -  \cos\frac{\theta}{2}    & -\sin\frac{\theta}{2}  e^{-\ii\phi}  \\
 - \sin\frac{\theta}{2}  e^{\ii\phi}   &   \cos\frac{\theta}{2} 
\end{pmatrix}= \mathrm{F}(\theta + \pi,\phi)= - \mathrm{F}(\theta,\phi)\,\mathrm{s}(\phi)\,. 
\end{equation}
We also note that the traceless features on both sides of \eqref{dotX} is respected. 

We now calculate the commutator $\left[\dot{U}U^{\dag}, \sigma_{\theta,\phi}\right]$ on the l.h.s. of \eqref{dotU}. From the relations
\begin{align}
\label{sigFs1}
\mathrm{F}(\theta,\phi)\, \sigma_{\theta,\phi}&= U(\pi-\theta,\phi)\, , \quad \sigma_{\theta,\phi}\, \mathrm{F}(\theta,\phi)= U^{\dag}(\pi-\theta,\phi)\,,\\
\label{sigFs2} 
\mathrm{s}(\phi)\,\sigma_{\theta,\phi}&=\sigma_{\theta + \frac{\pi}{2},\phi} = - \sigma_{\theta,\phi}\, \mathrm{s}(\phi)\Leftrightarrow \sigma_{\theta,\phi}\,\sigma_{\theta + \frac{\pi}{2},\phi}=-\mathrm{s}(\phi)= -\sigma_{\theta + \frac{\pi}{2},\phi}\,\sigma_{\theta,\phi}\,. 
\end{align}
we obtain 
\begin{equation}
\label{UUdsig}
\left[\dot{U}U^{\dag}, \sigma_{\theta,\phi}\right]= \ii \dot\phi\,\sin\theta\, \mathrm{s}(\phi) + \dot\theta\,\sigma_{\theta + \frac{\pi}{2},\phi} \,. 
\end{equation}
and
\begin{equation}
\label{UUdsigsig}
\left[\dot{U}U^{\dag}, \sigma_{\theta,\phi}\right]\sigma_{\theta,\phi}= \ii \dot\phi\,\sin\theta\, \sigma_{\theta + \frac{\pi}{2},\phi} + \dot\theta\,\mathrm{s}(\phi) \,. 
\end{equation}
The above set of  properties allow to obtain the separate equations for the angular variables. 
Starting from \eqref{dotUsig} we write
\begin{equation*}
\label{ }
\left[\dot{U}U^{\dag}, \sigma_{\theta,\phi}\right]\sigma_{\theta,\phi}= \ii \dot\phi\,\sin\theta\, \sigma_{\theta + \frac{\pi}{2},\phi} + \dot\theta\,\mathrm{s}(\phi)  = N(\theta,\phi)\,,
\end{equation*}
then multiplying each member of this equation by $\sigma_{\theta + \frac{\pi}{2},\phi}$ and tracing the result yield
\begin{equation}
\label{phidot}
\begin{split}
&2 \ii \dot\phi \, \sin\theta=\mathrm{Tr}\left[N(\theta,\phi)\,\sigma_{\theta + \frac{\pi}{2},\phi}\right]\\
&= 2\ii\left[ \sin\theta\,(h_{00}-h_{11}) - 2\cos\theta\,(\mathrm{Re}\,h_{01}\,\cos\phi - \mathrm{Im}\,h_{01}\,\sin\phi) + (h_2-h_1)\sin\theta\,\sin2\phi\right]\,,
\end{split}
\end{equation}
where the $h_{ij}$ are the matrix elements of $H$, $H=\begin{pmatrix}
   h_{00}   &  h_{01}  \\
  \overline{h_{01}}    & h_{11} 
\end{pmatrix}$.
After simplification, for $\theta\neq 0, \pi$, 
\begin{equation}
\label{phidot1}
\begin{split}
\dot\phi 
= h_{00}-h_{11} - 2\cot\theta\,(\mathrm{Re}\,h_{01}\,\cos\phi - \mathrm{Im}\,h_{01}\,\sin\phi) + (h_2-h_1)\,\sin2\phi\,.
\end{split}
\end{equation}
Similarly, multiplying \eqref{dotX} by $\mathrm{s}(\phi)$, tracing the result and dividing by 2 yield:
\begin{equation}
\label{thetadot1}
\dot\theta=  -2 (\mathrm{Re}\,h_{01}\,\sin\phi + \mathrm{Im}\,h_{01}\,\cos\phi)+ \sin2\theta(h_1\cos^2\phi + h_2\sin^2\phi -h_3)\,.
\end{equation}

These Ricatti-like equations can be interpreted as the equations of motion for the orientation of the Bloch vector on the sphere. They demonstrate how both the Hamiltonian and the decay rates contribute to the rotation and evolution of the quantum state's direction. The Hamiltonian terms $h_{ij}$ drive the coherent precession, while the decay rates $h_k$ induce a drift, or move towards a new non-rotational final position. The final state of the system is a balance between these two competing effects.

\subsubsection{The full dynamical system}

Collecting the evolution equations for angular and radial variables we finally obtain the system:
\begin{align}
\label{dsphi}
   \dot\phi 
&= h_{00}-h_{11} - 2\cot\theta\,(\mathrm{Re}\,h_{01}\,\cos\phi - \mathrm{Im}\,h_{01}\,\sin\phi) + (h_2-h_1)\,\sin2\phi\,,\\
\label{dstheta} \dot\theta &=  -2 (\mathrm{Re}\,h_{01}\,\sin\phi + \mathrm{Im}\,h_{01}\,\cos\phi)+ \sin2\theta(h_1\cos^2\phi + h_2\sin^2\phi -h_3)\, , \\
\frac{\dot{r}}{r} 
\label{dsr} &=-2\left[ h_1(1-\sin^2\theta\cos^2\phi)+h_2(1-\sin^2\theta\sin^2\phi) + h_3\sin^2\theta  \right]\,.   
\end{align}

As a simple example, for the particular cases where $H$ is diagonal and time-independent, and time independent $h_1,h_2, h_3$ with $h_1=h_2$, we obtain the solution 
\begin{align}
\label{simpsol1}
\phi(t) &= \phi_0 + \left(h_{00}-h_{11}\right)(t-t_0)\, ,\\
\label{simpsol2} \tan\theta(t)&=\tan\theta_0\,e^{-2(h_3-h_1) t} \,,\\
\label{simpsol3}  r(t)&=r_0\,e^{-4h_1 t}\,\vert\cos\theta_0\vert
\sqrt{1+\tan^2\!\theta_0\,e^{-4(h_3-h_1) t}} \,. 
\end{align}
Note the trajectories invariant which is derived  from \eqref{simpsol2} and \eqref{simpsol3}: 
\begin{equation}
\label{invariant}
r(t)\,|\cos\theta(t)|\,|\tan\theta(t)|^{-2h_1/(h_3-h_1)}
= r_0\,|\cos\theta_0|\,|\tan\theta_0|^{-2h_1/(h_3-h_1)}
\quad(h_3-h_1\neq 0)\,.
\end{equation} 
This is a geometric constraints equation that defines the exact shape of the spiral inside the Bloch ball.
The solution \eqref{eq: decayz} corresponds to the particular choices $h_1=0$, $h_3=h$, $t_0=0$, $\phi_0=0$, $h_{00}-h_{11}=2\omega_0$ , $r_0=1$, $\theta_0=\pi/4$.

\section{Fixed points: Discussion}
\label{fixed}
The asymptotic behavior of an open quantum system is characterized by its set of stationary or fixed points. As already presented in Subsec. \ref{subsec: onedirection}, the z-axis of the Bloch ball represent a set of fixed points. This corresponds to $h_1=h_2=0$, $\theta_0=0$, and $e_1=e_2=0$ ( in Eq. \eqref{eq: HamilCartesian} ), or $\mathrm{Re}\,h_{01}=\mathrm{Im}\,h_{01}=0$ ( in Eq. \eqref{dstheta} ). The other two axes (x and y) of the Bloch ball also are a set of fixed points, and this situation corresponds to the values of parameters $e_2=e_3=h_2=h_3=0$ for the x-axis ($\phi_0=0$, $\theta_0=\pi/2$) and $e_1=e_3=h_1=h_3=0$ for the y-axis ($\phi_0=\pi/2$, $\theta_0=\pi/2$). This can be verified by substituting those values in the equations \eqref{dsr}, \eqref{dstheta}, and \eqref{dsphi}. 

The fixed points of the system, denoted by $\vec{a}_*$, are identified by solving the steady-state condition $\vec{\dot{a}}=0$ for the governing dynamics in Eq.~\eqref{eq: CartesianEvo}. Once the fixed points are determined, their stability is ascertained through linear stability analysis. This procedure involves analyzing the evolution of an infinitesimal perturbation, $\vec{\delta}$, from a fixed point, such that $\vec{a}(t) = \vec{a}_* + \vec{\delta}(t)$. Substituting this into the  equations for the system and retaining only first-order terms in $\vec{\delta}$ yield a set of linear differential equations describing the dynamics of the perturbation. The stability of the fixed point is then determined by the eigenvalues of the corresponding Jacobian matrix.

To illustrate this methodology, we apply it to the system presented in Subsec.~\ref{subsec: onedirection}, whose dynamics are given by Eq.~\eqref{eq: dynsys1qbit}. The linearization around a fixed point $\vec{a}_*$ results in a system for the perturbation $\vec{\delta}$. By the definition of a fixed point, the static terms vanish, yielding the following linear system:
\begin{equation}
\label{eq:linearized_system}
\dot{\vec{\delta}} = -2\begin{pmatrix}
h & \omega_0 & 0 \\
-\omega_0 & h & 0 \\
0 & 0 & 0
\end{pmatrix}
\vec{\delta} \ .
\end{equation}
The eigenvalues of the Jacobian matrix are readily found to be $\lambda_1 = -2h - 2i\omega_0$, $\lambda_2 = -2h + 2i\omega_0$, and $\lambda_3 = 0$.

Because $h > 0$, the real parts of the complex conjugate eigenvalues $\lambda_{1,2}$ are negative. This implies that the dynamics are asymptotically stable in the $a_1a_2$-plane. Any perturbation in this plane will decay, returning the system to the fixed point. Conversely, the null eigenvalue, $\lambda_3=0$, indicates neutral stability along the $a_3$ direction. This signifies the existence of a continuum of equilibrium points. Collectively, these results demonstrate that the system possesses a line of fixed points, where each point on this line is an attractor for the dynamics within its respective $a_1a_2$-plane (i.e., for a constant $a_3$). Since the system under consideration is linear, this stability analysis is exact and provides a complete characterization of the equilibrium structure.

It is known \cite{SchirmerWang2009} that the existence of multiple steady states is fundamentally linked to the presence of symmetries or a decomposable structure in the  dynamics of the system. A system is considered decomposable if its Hilbert space can be broken down into two or more orthogonal subspaces that are invariant under the evolution. If such subspaces exist, the system cannot have a single globally attractive fixed point.

The study of fixed points and steady states is related to reservoir engineering\cite{Rakovszky}. By carefully designing a qubit's Hamiltonian and its dissipative interactions with an environment, one can control the final steady state the system evolves towards. This is a powerful tool for both state preparation, where a qubit is forced into a desired pure or mixed state, and for stabilization. In the latter case, the dissipative process actively protects a specific quantum state from noise by providing a form of continuous error correction, always pulling the state back towards the stable fixed point. Then, if the dynamics allow for multiple stable fixed points, this collection of states can be used as a robust memory. Each distinct fixed point can encode a piece of information, and the stability of these points makes the memory fault-tolerant or self-correcting, as the system's natural evolution will resist perturbations that try to alter the stored information.

\section{Conclusion}
\label{conclu}

In this paper, we have examined two complementary approaches to the GKSL equation for an open qubit. The first, essentially the standard method, exploits the linear structure of the equation and enables the systematic construction of explicit solutions; we illustrated it by analyzing representative trajectories within the Bloch ball.

 The second approach leverages the underlying SU$(2)$ symmetry of the Bloch ball and recasts the problem as a nonlinear dynamical system, thereby offering a geometrically motivated alternative description of the same open-system dynamics.

Although more intricate, this second formulation highlights in a natural way the distinction between the angular dynamics of the system and the radial component associated with dissipation. The main appeal of this second perspective lies in its potential for generalisation to the GKSL equation for open qudits, i.e., higher-dimensional generalisations of qubits.

Along these lines, the SU$(2)$–symmetry approach admits a natural generalisation by invoking the coset symmetry $\mathrm{SU}(N)/\mathrm{U}(1)^{N-1}$ together with a corresponding generalised angular-coordinate parametrisation. This extension is justified because, for an arbitrary $N$-level system (qudit), the Lie algebra $\mathfrak{su}(N)$ possesses $N-1$ independent Casimir invariants, which can be associated with a set of $N-1$ ``radial'' variables. As a consequence, one can partially decouple the dynamics of the GKSL equation: the evolution separates into contributions controlling the state's purity, the decay rates along the different Casimir directions, and the nonlinear dynamics governing the orientation of the generalised Bloch vector on the coset manifold. This provides a geometric framework in which dissipation and rotation are cleanly distinguished on the level of the dynamical system.

On the other hand, the linear analysis provided clear physical intuition, as illustrated by the case where dissipation affects only the $xy$-plane components, revealing a line of stable fixed points along the $z$-axis that acts as an attractor for the quantum states. These lines (the three axes) of stable fixed points can be employed to protect quantum information by engineering the right qubit-environment interactions.

\section{Aknownledgements}

During the preparation of this manuscript/study, the author(s) used Gemini and ChatGPT for the purposes of assist the writing. The authors have reviewed and edited the output and take full responsibility for the content of this publication. EMFC acknowledge the Conselho
Nacional de Desenvolvimento
Científico e Tecnológico - CNPq, and
the Fundação Carlos Chagas Filho
de Amparo à Pesquisa do Estado
do Rio de Janeiro - FAPERJ.

We thank the financial support from the Brazilian scientific agencies Funda\c{c}\~{a}o de Amparo \`a Pesquisa do Estado do Rio de Janeiro (FAPERJ), Coordena\c{c}\~{a}o de Aperfei\c{c}oamento de Pessoal de N\'ivel Superior (CAPES) and Conselho Nacional de Desenvolvimento Cient\'ifico e Tecnol\'ogico (CNPq).
T.K. acknowledges the financial support by CNPq (No.\ 305654/2021-7;304504/2024-6). 
A part of this work has been done under the project INCT-Nuclear Physics and Applications (No.\ 464898/2014-5)

\appendix
\section{Behavior for long times}\label{ap: longtimes}
In the sake of clarity, we here keep the expression $\sqrt{\beta^2-1}$ instead of using $\widetilde{\gamma}$. Following the example related to equation \eqref{eq: soludifferentregimes}, we can substitute the hyperbolic trigonometric functions with their exponential counterparts
\begin{align}
    \vec{a}(t)= & e^{-ht}\left(\left[\frac{e^{(2\omega_0t\sqrt{\beta^2-1})}+e^{-(2\omega_0t\sqrt{\beta^2-1})}}{2}+\frac{\beta \left(e^{(2\omega_0t\sqrt{\beta^2-1})}-e^{-(2\omega_0t\sqrt{\beta^2-1})}\right)}{2\sqrt{\beta^2-1}}\right], \right. \\ & \left.   \frac{e^{(2\omega_0t\sqrt{\beta^2-1})}-e^{-(2\omega_0t\sqrt{\beta^2-1})}}{2\sqrt{\beta^2-1}} ,\ 0\right)^T,
\end{align}
then, there are two exponentials with arguments $-t(h+2\omega_0\sqrt{\beta^2-1})$ and $-t(h-2\omega_0\sqrt{\beta^2-1})$, where the first decays faster then the other. So, for a long time
\begin{equation}
    e^{-ht-2\omega_0t\sqrt{\beta^2-1}}\rightarrow 0,
\end{equation}
and the system asymptotically behaves as
\begin{equation}
    \vec{a}(t)\rightarrow \left(\left[\frac{e^{-ht+2\omega_0t\sqrt{\beta^2-1}}}{2}+\frac{\beta e^{-ht+2\omega_0t\sqrt{\beta^2-1}}}{2\sqrt{\beta^2-1}}\right], \ \frac{e^{-ht+2\omega_0t\sqrt{\beta^2-1}}}{2\sqrt{\beta^2-1}} ,\ 0\right)^T.
\end{equation}
Then, we can define,
\begin{equation}
    \tau_d=\frac{1}{h-2\omega_0\sqrt{\beta^2-1}},
\end{equation}
after some manipulations it becomes the equation \eqref{eq: TimeDecay}. Thus, factoring out the exponentials, the norm is,
\begin{equation}
     \Vert \vec{a}(t)\Vert \rightarrow e^{-t/\tau_d}\left(\frac{\beta^2+\beta\sqrt{\beta^2-1}}{\sqrt{\beta^2-1}}\right)^{1/2} = e^{-t/\tau_d}\left(\frac{2\omega_0\tau_d\beta}{\sqrt{\beta^2-1}}\right)^{1/2}.
\end{equation}
Therefore, for long times $t\gg\tau_d$
\begin{equation}
    \Vert \vec{a}(t)\Vert \propto e^{-t/\tau_d}.
\end{equation}

\section[\appendixname~\thesection]{One qbit dynamic system}\label{ap: DynSys}
The GKSL equation \ref{eq: lindbladA} with density operators
\begin{equation}\label{eq: lindbladrho}
    \frac{\mathrm{d} \rho}{\mathrm{d} t}=\ii\,[H,\rho]+\sum _{k=1}^3h_k\left[\sigma_k\rho \sigma_k -\rho\right] \, ,
\end{equation}
where the jump operators the Pauli matrices substituted $L_{k}$, becomes easily solvable by writing the density matrix with Einstein's notation
\begin{equation}
    \rho=\frac{1}{2}\left(\mathbbm{1}_2+a_1 \sigma_1+a_2 \sigma_2+a_3 \sigma_3\right)\equiv \frac{1}{2}\left(\mathbbm{1}_2+a_j \sigma_j\right),
\end{equation}
and analogously, the Hamiltonian becomes
\begin{equation}
    H=e_0\mathbbm{1}_2 +e_\mu\sigma_\mu \ ,
\end{equation}
with $j,\mu=1,2,3$. Then, the commutator of equation \ref{eq: lindbladrho} is written as
\begin{align}
    \ii[H,\rho]= & \frac{\ii}{2} \left[\left(e_0\mathbbm{1}_2 +e_\mu\sigma_\mu\right)\left(\mathbbm{1}_2+a_j \sigma_j\right)-\left(\mathbbm{1}_2+a_j \sigma_j\right)\left(e_0\mathbbm{1}_2 +e_\mu\sigma_\mu\right)\right], \\
    = & \frac{\ii}{2} \left[e_\mu\sigma_\mu a_j \sigma_j- a_j \sigma_je_\mu\sigma_\mu\right], \\
    = & \frac{\ii e_\mu a_j}{2}\left(\ii \epsilon_{\mu j l}-\ii \epsilon_{j\mu  l}\right) =-e_\mu a_j \epsilon_{\mu j l} \sigma_l \ , \label{eq: comut}
\end{align}
while the derivative
\begin{equation}\label{eq: derivative}
    \frac{\ud\rho}{\ud t}=\frac{\dot{a}_m}{2}\sigma_m \ .
\end{equation}
Before determining directly the sum in equation \ref{eq: lindbladrho}, it is important to state that any matrices of order $2$ can be written in terms of the identity and the Pauli matrices, for example, the matrix $M$
\begin{equation}
    M=m_0\mathbbm{1}_2+m_1\sigma_1+m_2\sigma_2+m_3\sigma_3,
\end{equation}
and the coefficients $m_0$, $m_n$ can be obtained by
\begin{align}
    m_0= & \frac{1}{2}{\rm Tr}\left[M\right], \\
    m_n= & \frac{1}{2}{\rm Tr}\left[\sigma_n M\right] \ .
\end{align}
Thus, the coefficients of the left-hand side of equation \eqref{eq: lindbladrho} can be extracted from equation \eqref{eq: derivative}
\begin{equation} \label{eq: derivcomp}
    \frac{1}{2}{\rm Tr}\left[\sigma_n\frac{\ud\rho}{\ud t}\right]= \frac{\dot{a}_n}{2},
\end{equation}
and the $n$-th coefficient of the commutator in equation \eqref{eq: comut}
\begin{equation}\label{eq: comut2}
    \frac{1}{2}{\rm Tr}\left[\sigma_n \ \ii\left[H,\rho\right] \ \right]= -e_\mu a_j \epsilon_{\mu j n} \ .
\end{equation}
Now, one can determine the coefficients related to the sum of equation \eqref{eq: lindbladrho}
\begin{equation} \label{eq: lcof}
    \frac{1}{2}{\rm Tr}\left[\sigma_n\left(\sum_kh_k\sigma_k\rho\sigma_k-h_k\rho\right) \right]=\sum_{k,l=1}^3h_k\left\{\frac{a_l}{4}{\rm Tr}\left[\sigma_n\sigma_k\sigma_l\sigma_k\right]-\frac{a_n}{2}\right\} \,
\end{equation}
with the help of the identity ${\rm Tr}\left[\sigma_i\sigma_j\sigma_k\sigma_l\right]=2\left(\delta_{ij}\delta_{kl}-\delta_{ik}\delta_{jl}+\delta_{il}\delta_{jk}\right)$, the equation \eqref{eq: lcof} can be written as
\begin{equation}\label{eq: somalindblafin}
     \frac{1}{2}{\rm Tr}\left[\sigma_n\left(\sum_kh_k\sigma_k\rho\sigma_k-h_k\rho\right) \right]=h_na_n-a_n\sum_{k=1}^3h_k \ .
\end{equation}
Therefore, we can rewrite the GKSL equation based on equations \eqref{eq: derivcomp}, \eqref{eq: comut2}, and \eqref{eq: somalindblafin} 
\begin{equation}
    \frac{\dot{a}_n}{2}=-e_\mu a_j \epsilon_{\mu j n}+a_n\left(h_n-\sum_{k=1}^3 h_k \right),
\end{equation}
which gives the set of equations \eqref{eq: dynsys1qbit}.

\end{document}